\documentclass[aps,prl,twocolumn,nofootinbib,superscriptaddress,showpacs]{revtex4-1}
\usepackage{graphicx}
\usepackage{dcolumn}
\usepackage{bm}

\def\pc{\rm pc}

\def\msun{M_\odot}
\def\rsun{R_\odot}
\def\ten#1{\times 10^{#1}}
\def\aten#1{10^{#1}}
\def\VEV#1{\left\langle #1\right\rangle}

\def\mbh{m_{\rm PBH}}
\def\avete{\VEV{t_e}}
\def\mn{{m_{9}}}
\def\amax{A_{\rm max}}
\def\ustarmax{u_{*{\rm max}}}
\def\xmax{{x_{\rm max}}}
\def\dgdte{{d\Gamma\over dt_e}}
\def\betaprime{{\beta'}}
\def\uthresh{u_{\rm thresh}}
\def\los{{\it los}}

\begin{document}


\title{Microlensing of Kepler Stars as a Method of Detecting Primordial Black Hole Dark Matter}


\author{Kim Griest}
\affiliation{Department of Physics, University of California, San Diego, CA 92093, USA.}

\author{Matthew J. Lehner}
\affiliation{Institute of Astronomy and Astrophysics,
    Academia Sinica. P.O. Box 23-141, Taipei 106, Taiwan}
\affiliation{Department of Physics and Astronomy,
    University of Pennsylvania, Philadelphia, PA 19104.}

\author{Agnieszka M. Cieplak}
\affiliation{Department of Physics, University of California, San Diego, CA 92093, USA.}

\author{Bhuvnesh Jain}
\affiliation{Department of Physics and Astronomy,
    University of Pennsylvania, Philadelphia, PA 19104.}

\email{\tt kgriest@ucsd.edu, acieplak@ucsd.edu, mlehner@asiaa.sinica.edu.tw}


\date{\today}


\begin{abstract}
If the Dark Matter consists of primordial black holes (PBHs),
we show that gravitational lensing of stars being monitored by NASA's
Kepler search for extra-solar
planets can cause significant numbers of detectable microlensing events.
A search through the roughly 150,000 lightcurves would result in large
numbers of detectable events for
PBHs in the mass range $5 \ten{-10}\msun$ to $\aten{-4}\msun$.
Non-detection of these events would close almost two orders of
magnitude of the mass window for PBH dark matter.
The microlensing rate is higher than previously noticed due to a
combination of the
exceptional photometric precision of the Kepler mission and the
increase in cross section due to the large angular sizes of the
relatively nearby Kepler field stars.  We also present a new formalism
for calculating optical depth and microlensing rates in the presence of
large finite-source effects.
\bigskip

\end{abstract}


\maketitle

\section{Primordial Black Hole Dark Matter}
Primordial Black Holes (PBHs) have been considered as a candidate for dark matter (DM)
since the days of Hawking\cite{hawking74,hawking75} and PBHs are recently 
re-emerging as objects of intense
study (see references\cite{carr,josan} for recent reviews of constraints).
PBH DM can form during the early universe as a result of 
large density fluctuations coming
from, for example, bubble collisions, 
collapse of string loops,
features in inflation potentials, etc. (see reference\cite{frampton}
for more examples and references). 
If PBHs form early enough they can evade the Big Bang Nucleosynthesis
limits on baryons,
satisfy CMB constraints, and make up the entirety of the dark matter.

While there is no compelling mass or mass range for PBH DM, there has been
extensive experimental and theoretical work that
has eliminated most mass
ranges, starting with Hawking's famous limit, 
$\mbh> \aten{15}$g ($\mbh>\aten{-18}\msun$) 
for PBHs to not have evaporated by today. 
See reference\cite{carr} for a recent summary of these constraints. 
Currently 
most mass ranges from $\aten{-18}\msun$
to $\aten{16}\msun$ are ruled out, 
with the exception being the 
five orders of magnitude between 
$\aten{-13}\msun < \mbh <\aten{-7}\msun$, which we call the PBH
DM window.
As we show below, 
microlensing of Kepler source stars 
can detect or rule out PBH DM over a large fraction of the PBH DM
window.  Thus, if the DM consists of PBHs, the experiment we propose here has
an excellent chance of detecting it.  Since gravitational 
microlensing is sensitive to any massive compact halo object (MACHO) 
a detection or limit would actually be more general than just PBH DM.

\section{Kepler Mission}
The NASA Kepler satellite is a 1 m aperture telescope with a 
115 deg$^2$ field-of-view in an Earth
trailing heliocentric orbit, which is currently taking photometric 
measurements of around 150,000 stars every 30 
minutes\cite{boruckiscience,archivemanual,keplerwebsite}.
The telescope was launched in
March 2009 and will point at the same field (in the Cygnus-Lyra region of 
the sky) for at least 3.5 years.  
The goal of the Kepler mission is to find extra-solar planets via small
decreases in stellar flux due to rare planetary transits in front of 
the host star in edge-on systems.
Very precise photometry is required to detect the tiny decrease in flux
that an Earth size planet causes during a transit.

While much of the data
is still proprietary, in what follows
we analyze a small portion of it to investigate its
potential use in a microlensing experiment \cite{mast}.  
Looking through a portion of this
publicly available data we see that Kepler source stars have 
V magnitudes between
roughly 10 and 16, have distances from the Earth between 0.9 and 3 kpc 
(with most between 0.9 and 1.6 kpc), have stellar radii, 
$R_*$, between $0.9\rsun$ and $1.5\rsun$ 
(for main sequence stars, but there are substantial
numbers of giant stars as well with $5\rsun < R_* < 20\rsun$), 
and have photometric errors per observation between 
20ppm and 1000ppm (with most between
300pm and 1000pm).

\section{Microlensing and Analytic Estimate}
Microlensing searches for low mass objects have been performed and 
have returned significant limits on any massive compact halo objects, 
including PBHs, 
that might constitute the dark matter\cite{machospike, erosmacho, eros, ogle}.

These microlensing surveys examined tens of millions stars in 
the Large Magellanic Cloud (LMC), which is at a distance of 50 kpc, 
for periods of many years.
Since the naive microlensing optical depth increases with source star 
distance and the total
microlensing event rate is proportional to the number of monitored stars 
times the duration of the survey,
it might be surprising that the many fewer Kepler lightcurves on much closer
stars, can provide stronger limits on low mass DM in 
certain mass ranges than these surveys have.  

As calculated below, this is due to several factors.   
First, the extreme precision of the Kepler photometry allows very small
magnifications to be detected.  
Requiring a sequence of 2-sigma or 3-sigma measurements on a typical
Kepler lightcurve allows a magnification threshold of $A_T = 1.001$ or 
lower to be set (magnification, $A \equiv ({\rm Flux}/\VEV{\rm Flux}) -1$).
Since detection is said to occur whenever the magnification 
$A>A_T$, the effective microlensing tube\cite{griest91} is wider by 
a factor of $u_T = [2A_T(A_T^2-1)^{-1/2} - 2]^{1/2} \sim 6$.

Here we are using notation\cite{pac,griest91} 
in which the Einstein ring radius 
(radius of image ring arising from perfectly aligned source and lens) is
$r_E=0.0193 \sqrt{x(1-x)}[(D_s/{\rm kpc})\mn]^{1/2}\rsun$, where
$\mn=(m/\aten{-9}\msun)$, is the mass of the PBH DM lens, 
$D_s$ is the source star distance, and
$x= D_l/D_s$ is the ratio of the lens distance to source distance. 
We define the standard microlensing ``tube" as that volume along 
the line-of-sight inside the Einstein ring, and define $u$ to be the transverse
distance between the lens and the source center in units of $r_E$.
The important parameter, $u_T$, is the ``effective" radius of the microlensing
tube in units of $r_E$ in the point source limit; thus the usual standard 
$u_T=1$ implies $A_T=1.34$.

The usual point source (PS) 
optical depth (total number of PBHs inside the microlensing tube),
$\tau_{\rm PS} = \int_0^1 dx D_s\rho(x)\pi u_T^2 r_E^2/m$, 
is thus larger than in the standard case by a factor of $u_T^2$. 
where
$\rho(x)$ is the run of DM density along the line-of-sight (\los).

However, just as important for low mass PBH microlensing
is the fact that the effective microlensing tube radius is not determined
by the Einstein ring radius, but by the radius of the source star due
to finite-source effects.
For very low mass
PBHs, the Einstein ring radius is much smaller than the 
projected source star radius, $x R_*$,
and $u_T$ in the formula for optical 
depth should be replaced by $\uthresh$ which is the value of
the source-lens transverse separation that gives $A=A_T$.
The quantity $\uthresh$ may be approximated by $u_* = xR_*/r_E$, though
below we make a better approximation using the finite-source lightcurve
formulas from Witt \& Mao\cite{witt}.
For $m \approx \aten{-9} \msun$ with 
$x \approx 0.5$ and a typical Kepler source 
star with $R_*=1\rsun$ and 
$D_s=1$ kpc, we find $u_*=51.9$, which gives an 
optical depth (and total event rate)
around 2,700 times larger than naively expected.  
(See below for a more accurate estimate.) 
Similarly, the usual statement that the optical depth is 
independent of DM mass is not true in this case.
Naively, a lower mass means a narrower microlensing tube, 
but more DM objects, effects that
cancel each other out in the optical depth.  
For the case where all events are finite-source
events, lowering the mass does not change the tube radius, 
so the optical depth is roughly
inversely proportional to mass, 
increasing the sensitivity to much lower mass DM objects.
Likewise, the average event duration, 
$\avete$,  defined as the time for which the source star 
is magnified above $A_T$, is now roughly independent of mass, and
the total event rate, $\Gamma = \tau/\avete$, now 
increases roughly inversely with decreasing mass.

Of course, finite-source microlensing has been studied since the paper
by Witt \& Mao\cite{witt},
who show that in the finite source regime
the magnification is limited to be less than
$\amax = \sqrt{1 + 4/u_*^2}$.  For low mass PBHs, the lightcurve jumps
quickly from near $A=1$ to $A=\amax$ and then 
stays roughly constant (actually following the limb-darkening shape of 
the source star) as the small Einstein ring 
transits the stellar limb, jumping back to near $A=1$ at the edge.
See references\cite{witt,machospike} for example lightcurves,
which are quite different from standard microlensing lightcurves.
The very precise Kepler photometry is what allows
these low magnification finite-source lightcurves to be detected.   
For a given $A_T$, we use the previous equation to find a cutoff, 
$\ustarmax$, such that a microlensing event is observable whenever $u$
is less than $\ustarmax = 2/\sqrt{A_T^2-1}$.

The total optical depth and the total event rate include integrals along 
the \los, but the limits to these integrals need modification when
considering strong finite-source effects.
As $x\rightarrow 0$, both the projected source star 
radius, $xR_*$, and $r_E$, go to zero. 
Their ratio, which is $u_*$, also goes to zero as $x\rightarrow 0$, while
as $x\rightarrow 1$, $r_E\rightarrow 0$ and  $u_*\rightarrow \infty$.  
Because of this last dependence, $\amax\rightarrow 1$ as $x\rightarrow 1$,
and so there is always some limiting value of distance,
$\xmax$, where only objects that enter the tube with $x<\xmax$ give 
detectable magnifications.  We find
$\xmax = 1/(1+\kappa^2)$, where $\kappa= 51.9 (R_*/\rsun) 
/(\ustarmax \sqrt{\mn D_s/{\rm kpc}})$.
Thus the optical depth integration limits above should be 
between $x=0$ and $x=\xmax$, not between 0 and 1.  

For the pure finite source case ($u_*\gg u_T$) and a constant DM density,
the above optical depth integral can be easily performed.
We find $\tau_{\rm FS} \approx {\pi \over3} D_s \rho R_*^2 \xmax^3/m 
 \approx 4.2\ten{-6} \xmax^3 (R_*/\rsun)^2 (D_s/{\rm kpc})/\mn$.
For $A_T=1.001$, $R_* = 1 \rsun$, $D_s=1$ kpc, 
and $\mn=1$, we have $\xmax=0.426$ and $\tau_{\rm FS}=3.2\ten{-7}$.
To estimate the total event rate in this case we can use the very
approximate formula from reference\cite{pac} or \cite{griest91}, 
replacing $u_T r_E$ with $\xmax R_*$.  We find
$\Gamma_{\rm Pac-FS} \approx 2\tau_{\rm FS} v_c/(\pi \xmax R_*) = 
4.8\ten{-3}$ events/year/star.
The corresponding average event duration is given by 
$\avete = \tau/\Gamma \approx 0.59$ hours.
This is a remarkably large event rate and if 150,000 stars 
were followed it would result
in a total of around 720 events per year.

However, another complication is that for a microlensing event to be 
detected, it must contain
enough significant measurements to be unlikely to occur by chance.  
For example, given Kepler's 30 minute cadence, 
monitoring 150,000 stars means about 2.6 billion
flux measurements per year.  Assuming Gaussian errors, the 
probability of finding 4 sequential measurements 3-sigma above average 
is such that less than one such instance will occur in  3.5 years of
data.  This requirement implies an event 
duration, $t_e$, the time for which $A>A_T$, of at least 2.0 hours. 
This means that we cannot use the total event rate $\Gamma$, but must 
integrate the differential event rate $\dgdte$
over a relevant range of event durations.  
This differential event rate is given
in Equation~17 of reference\cite{griest91} for non-finite-source events.  
In the limit of large $u_*$
this formula can be used just by replacing $u_T$ with 
$u_*$ throughout.  There are also
two simplifications to this formula due to the unique position of 
the Kepler field.
First, the Kepler field is at galactic longitude and latitude 
$(l,b)=(76.32^0,13.5^0)$ which places
it just out of the plane of the Milky Way disk, in almost the 
same direction in which the solar system
is moving\cite{archivemanual}.  
The Kepler source stars, at distances of 1 to 3 kpc, 
are thus at nearly the same distance
from the Milky Way (MW) center 
as the Sun (8.5 kpc).  
Thus, unlike the case for the LMC, there is no need to model
the halo density and one can just use the local dark matter density 
$\rho \approx 0.3\ {\rm GeV cm}^{-3} =7.9\ten{-3}\msun {\pc}^{-3}$ 
along the entire \los.
Second, since the Kepler stars are in the direction of solar motion and 
also orbiting the MW center, there is no additional large transverse velocity 
of the Sun or source to include.
Thus Equation~17 of reference\cite{griest91} becomes
\begin{equation}
\dgdte = {\rho\over m}D_s v_c^2 \int_0^\xmax dx \betaprime^2 g(\betaprime),
\label{eqn:eqndgdte}
\end{equation}
where $g(\betaprime) = \int_0^1 dy  
y^{3/2} (1-y)^{-1/2} e^{- \betaprime y} = {\pi\over 2}e^{-\betaprime/2} [
I_0(\betaprime/2) - (1+1/\betaprime)I_1(\betaprime/2)]$,
$I_0$ and $I_1$ are modified Bessel functions of the first kind,
$\betaprime = 4 r_E^2 \uthresh^2 /(t_e^2 v_c^2)$, $y=v_r^2/(\betaprime v_c^2)$,
$v_c \approx 220$ km/s, is the halo circular velocity, 
and we introduced $\uthresh$ which stands
for $u_T$ in the case of point source lensing, 
$u_*$ in the case of pure finite-source lensing,
and below will be approximated as something in between for the general case.
For use later we note that $g(0) = {3\over8}\pi$, and $g(\betaprime) \sim
{3\sqrt{\pi}\over 4} \betaprime^{-5/2}$ for large values of $\betaprime$.

\section{Numerical Estimate}
In order to make a reasonably accurate estimate of the potential sensitivity 
of a search through the Kepler lightcurve data we need to integrate 
Eq.~\ref{eqn:eqndgdte} from $t_e=t_{min}$ to some
reasonable upper limit.  We also want to use realistic distributions of 
Kepler star distances, radii, and $A_T$, and 
we want an approximation for $\uthresh$ 
that interpolates accurately between $\uthresh \approx u_T$ at $x$ near 0, and 
$\uthresh\approx u_*$ at $x$ near 1. 
Ideally, we would also include the effects of limb-darkening, but do not
do this here.  We have done preliminary exploration
of this effect by creating limb-darkened finite-source lightcurves
and calculating $\uthresh$ for these, but do not see a large effect 
for the few PBH mass values we checked.

We looked at a subsample of around 5000 publicly available Kepler lightcurves 
from the NASA MAST website\cite{mast}.
We used data from the third publicly available quarter and, 
besides the lightcurves of fluxes
and flux errors, we have for each star:
the stellar radius, $R_*$, 
Sloan $r$ and $g$ magnitudes, effective temperature, $T_\mathrm{eff}$, 
star position,
extinction parameters $A_V$ and $E(B-V)$, etc. 
We estimate the apparent visual magnitude, $V = g - 0.0026 - 0.533(g-r)$
\cite{fukugita}, and the stellar distance from
$D_s = 1.19\ten{-3} R_* (T_\mathrm{eff}/T_\odot)^2 
\aten{0.2(V-A_V+{\rm B.C.})}\mathrm{kpc}$,
where B.C. is the bolometric correction.  
Note that we make a very crude bolometric correction, using 
only the effective temperature and whether the source is a main sequence
or giant star\cite{carrollostlie}, but 
we include it because it slightly reduces the distances to the sources,
thereby reducing the expected detection rate, and we want
our calculation to be conservative.

To find $A_T$ for each source star,
we calculated the average of the reported flux errors over 300 data 
points near the middle of each lightcurve, and then estimated each 
$A_T$ as one plus three times this average (for
a 3-sigma detection requirement).  
We also calculated the standard deviation of the flux measurements
over this portion of the lightcurve and found reasonable agreement 
with the average flux error.  For our subsample of 5000 stars we
find $2\ten{-5} < A_T < 3\ten{-3}$, with vast majority having
$A_T\approx0.001$.

As noted above, the approximation, 
$\uthresh\approx u_*$ misses some events, since for nearby lenses
the projected source radius becomes smaller than the Einstein ring radius, and 
therefore $\uthresh$ should approach $u_T$ rather than zero.  
To develop a better approximation, we numerically calculated 
a large set of finite-source lightcurves using the formula 
from Witt and Mao\cite{witt}, and then for 
each lightcurve calculated the actual value of 
$\uthresh$ (value of lens-source transverse separation for which
$A>A_T$) as a function of $u_*/u_T$.  We find that $\uthresh/u_T$ is a fairly universal function
of $u_*/u_T$, and therefore we can calculate $u_*$ for each value of $x$ and find $\uthresh$ using this
universal function.  Our fit function is:
$\uthresh \approx u_T(1 + .47 (u_*/u_T)^2)$, for $u_*/u_T<0.75$, $\uthresh \approx u_*$, for
$u_*/u_T>4.5$, and $\uthresh \approx u_*/(\sum_i c_i (u_*/u_T)^i)$, with $c_0=0.0971$, $c_1=0.925$,
$c_2=-0.384$, $c_3=0.0723$, and $c_4 = -0.0051$, for $0.75 < u_*/u_T < 4.5$.  
This approximate
$\uthresh$ differs from our numerically calculated values by no more than 
2.5\% over the mass range we investigated.

We then numerically performed the 2-dimensional integral of Eq.~\ref{eqn:eqndgdte} for
various values of PBH mass, various S/N requirements (e.g. four 3-sigma measurements or 
seven 2-sigma measurements).  We summed the total number of expected
events over the 5000 stars and then scaled those results to 3.5 
years of observation of 150,000 stars, assuming that 25\% of these stars
will be identified as variable and not be useful\cite{keplerwebsite}.
That is we assume 390,000 star-years.
Our results are given in Figure~1.  In order to turn these results into 
the potential sensitivity of detecting PBH dark matter,
we calculated the 95\% C.L. for each PBH mass, assuming that no
events were detected and that there was no 
background.  These potential limits are shown in Figure~1,
along with limits from earlier experiments.
We found very little difference between the two S/N 
requirements mentioned above, so show only the case for 4 sequential 3-sigma
event selection.

We see from Figure~1 that we have the potential to detect or 
rule out PBHs as the primary
constituent of DM over the mass range $5\ten{-10}\msun < \mbh < \aten{-4}\msun$.
Current limits from the MACHO/EROS experiments\cite{erosmacho}
are also shown as a dashed line.
There are also limits ruling out halo fractions, $f>1$, from femtolensing 
of gamma ray bursts (GRB)\cite{marani},
but they run from about $\aten{-16}\msun < \mbh < \aten{-13}\msun$, 
off to the left 
of Figure~1, and other limits from picolensing of GRBs ruling out $f<4$ from 
$4\ten{-13}\msun < \mbh < 8\ten{-10}\msun$, 
are in our mass range, but too weak to be seen in our plot.
We see that a microlensing search for PBHs through Kepler data 
has the potential
to extend the mass sensitivity by almost two orders of magnitude below
the MACHO/EROS limits which exclude DM masses down to 
around $2\ten{-8}\msun$.  
Note that commonly quoted (e.g. \cite{carr}) limits from 
EROS alone\cite{eros} 
exclude masses down to around $6\ten{-8}\msun$ and
are not as strong as the earlier combined MACHO and EROS limits.
There are no other limits in the mass range just below $2\ten{-8}\msun$, 
so the capability of Kepler to search for these PBHs is unique.

\begin{figure}
\includegraphics[scale = 0.4]{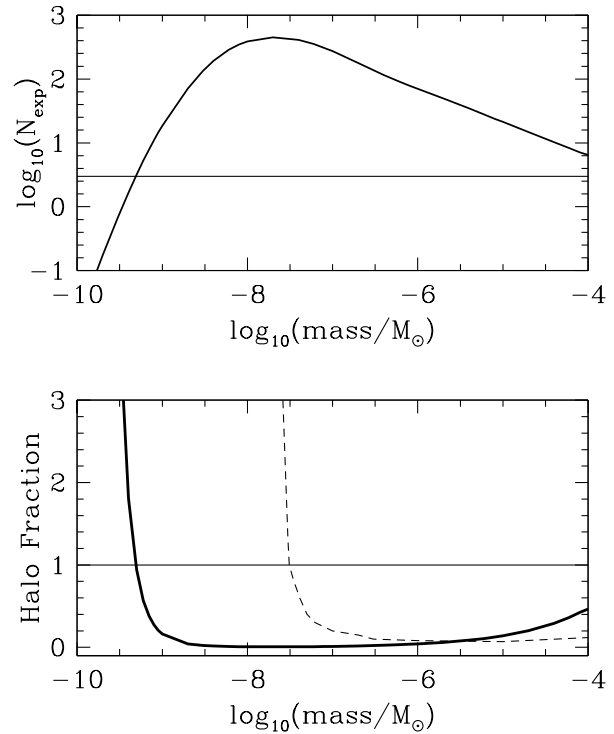}
\caption{
\label{figlimits}
{\it Top panel:} The expected number of events, $N_{\rm exp}$
(defined as 4 sequential measurements with flux 3-sigma above average) in
390,000 star-years of Kepler data. The thin horizontal line 
shows $N_{\rm exp}=3$, the limit for a 95\% C.L. if no events are detected.
{\it Bottom panel:}
Potential 95\% C.L. exclusion of PBH dark matter.
The area above the thick line would be ruled out if no events were seen
in 390,000 star-years. 
Also shown as a thin dashed line is the limit from combined 
MACHO/EROS LMC microlensing searches\cite{erosmacho}.
The thin horizontal line shows a 
halo consisting entirely of PBHs with local
DM density $\rho = 0.3 {\rm GeV cm}^{-3}$.}
\end{figure}

\section{Discussion and Future Work}
In this theoretical paper we suggest a method to detect or rule out 
PBH DM over a large and unexplored mass range.  If nothing is detected
the method has the potential to rule out around 40\% of the total remaining
mass range for PBH DM.
While we emphasized PBHs, 
since microlensing depends only upon lens mass 
and size, such an
experiment would detect or rule out any massive compact 
halo object dark matter in the mass range described.

The next step is clearly to perform the analysis suggested here,
and we have begun this task using publicly available Kepler data.
This experiment requires a deep understanding of the 
lightcurve data including systematic effects
and possible backgrounds.  
It requires a careful selection of microlensing candidates
and an accurate calculation of the efficiency of that selection method, 
as well as an understanding of false positives
and background events.  
Since Kepler transit searches have already turned up many
stellar flares on M and K dwarfs\cite{walkowicz}
using selection criteria similar to those
used above, a good method of distinguishing microlensing from stellar 
flares will be needed.
However, there are some powerful discriminants against 
backgrounds.
The lightcurves should match a limb-darkened finite source 
microlensing shape, and if enough events are found the rates should exhibit
the strong dependence on distance and source radius predicted by the above
formulas.  

Several effects in the data may mean that 
the actual limits, supposing no
PBHs are detected, will be different from what we display 
in Figure~1.  First, our simplistic removal of 25\% of the stars as
variable\cite{keplerwebsite} needs to be replaced by detailed analysis of 
all the lightcurves.  Our calculation of thresholds using Gaussian statistics
needs redoing in light of actual lightcurve data and other sources 
of noise that exist in real data.
A more careful treatment
of stellar limb darkening may shorten the average duration of events.
The actual selection of candidate events may turn-up poor 
S/N microlensing candidates that will reduce the significance of any 
limits derived.  These effects will reduce the expected number of events 
used to calculate Figure~1.  Of course, if the Kepler mission
runs longer than 3.5 years
the sensitivity would be better than that in Figure~1, and it also
might be possible to use lightcurve shape discrimination to detect 
lower magnification events, thereby increasing sensitivity.

There are also several assumptions and simplifications
in our analysis that could change the results.
For example our estimate of the distances to the source stars is
overly simple, as is our treatment of limb-darkening.
We ignored the MW halo model and any transverse velocity 
of the Sun and Kepler source stars.  
Since the Kepler field is in the direction of the Sun's orbit about the MW, 
any rotation of the DM halo in the direction of the disk would give 
support to the halo and not contribute to transverse microlensing velocity.  
This would mean that the average transverse velocity of the PBHs
relevant for microlensing might be lower, 
implying the average duration of events might be longer.  
Longer events are easier to detect, so if this is the case, our 
limits might be conservative.

\begin{acknowledgments}
\acknowledgments
K.G. thanks the IPMU (Tokyo) Focus Week on Dark Matter, where some of 
the ideas for this paper were started.

K.G. and A.M.C. were supported in part by 
the DoE under grant DE-FG03-97ER40546.  
A.M.C. was supported in part by the National Science Foundation Graduate
Research Fellowship under Grant Number DGE0707423.
B.J. is supported in part by NSF grant AST-0908027 and DOE grant 
DE-FG02-95ER40893.

Some of the data presented in this paper were obtained from the 
Multimission Archive at the Space Telescope Science Institute (MAST). 
STScI is operated by the Association of Universities for Research in 
Astronomy, Inc., under NASA contract NAS5-26555. Support for MAST for 
non-HST data is provided by the NASA Office of Space Science via grant 
NNX09AF08G and by other grants and contracts.

\end{acknowledgments}

%

\end{document}